\documentclass[11pt]{article}
\usepackage{mypre}
\usepackage{epsfig}

\begin{document}

\title{\bf\boldmath${\cal F}(1)$ for $B\to D^*l\nu$ from Lattice
QCD\thanks{presented by A.~S.~Kronfeld at {\em Flavor
Physics and $CP$ Violation}, May 16--18, Philadelphia, Pennsylvania}}

\author{Andreas S. Kronfeld, Paul~B. Mackenzie, James~N. Simone \\
{\em\small
Fermi National Accelerator Laboratory, 
Batavia, Illinois 60510, USA}
\and
Shoji Hashimoto \\
{\em\small High Energy Accelerator Research Organization (KEK)
Tsukuba~305-0801, Japan}
\and
Sin\'ead M. Ryan \\
{\em\small School of Mathematics, Trinity College, Dublin~2, Ireland}}
\date{July 9, 2002}     

\preprint{\sf FERMILAB-Conf-02/145-T \\ \tt hep-ph/0207122}

\maketitle

We would like to determine $|V_{cb}|$ from the exclusive semi-leptonic
decay $B\to D^*l\nu$.
The differential decay rate is
\begin{equation}
	\frac{d\Gamma}{dw} = \frac{G_F^2}{4\pi^3} (w^2-1)^{1/2}
	m_{D^*}^3 (m_B-m_{D^*})^2 {\cal G}(w)
	|V_{cb}|^2 |{\cal F}_{B\to D^*}(w)|^2,
\end{equation}
where $w=v\cdot v'$ and ${\cal G}(1)=1$.
At zero recoil ($w=1$) heavy-quark symmetry requires
${\cal F}_{B\to D^*}(1)$ to be close to~1.
So, $|V_{cb}|$ is determined by dividing measurements of $d\Gamma/dw$
by the phase space and well-known factors, and extrapolating
to $w\to 1$.
This yields $|V_{cb}|{\cal F}_{B\to D^*}(1)$,
and ${\cal F}_{B\to D^*}(1)$ is taken from ``theory.''
To date models~\cite{Neubert:1994vy} or a combination of a rigorous
inequality plus judgment~\cite{Shifman:1995jh}
have been used to estimate ${\cal F}_{B\to D^*}(1)-1$.
In this work~\cite{Hashimoto:2001nb} we calculate 
${\cal F}_{B\to D^*}(1)$ with lattice gauge theory, in the so-called 
quenched approximation, but the uncertainty from quenching is included 
in the error budget.

The ``form factor'' ${\cal F}_{B\to D^*}(w)$ is a linear combination of
several form factors of the matrix elements
$\langle D^*|{\cal V}^\mu|B\rangle$ and
$\langle D^*|{\cal A}^\mu|B\rangle$.
At zero recoil all form factors but $h_{A_1}$ are suppressed by phase 
space, so
\begin{equation}
	{\cal F}_{B\to D^*}(1) = h_{A_1}(1) =
		\langle D^*(v) | {\cal A}^\mu | B(v) \rangle,
\end{equation}
which should be ``straightforward'' to calculate in lattice QCD.
But a brute force calculation of $\langle D^*|{\cal A}^\mu|B\rangle$
would not be interesting: similar matrix elements like
$\langle 0|{\cal A}^\mu|B\rangle$ and 
$\langle\pi|{\cal V}^\mu|B\rangle$ have 15--20\% 
errors~\cite{El-Khadra:1998hq,Khadra:2001ti}.

Thus, we have to involve heavy-quark symmetry from the outset: if we
can focus on $h_{A_1}-1$, we have a chance of success, because a 20\%
error on $h_{A_1}-1$ is interesting: $0.2\times0.1=0.02$.
There are three specific obstacles to overcome:
(i)~statistical uncertainties,
(ii)~normalization uncertainties in the lattice axial vector current,
and (iii) how to treat heavy quarks since $m_ba\not\ll1$.
The first two need computational insight; 
the last two theoretical insight.
In the last several years, we have developed tools to attack these 
problems~\cite{Lepage:1993xa,El-Khadra:1997mp,Kronfeld:2000ck,%
Harada:2001fi,Harada:2001fj,Hashimoto:2000yp}.

At zero recoil heavy-quark symmetry 
implies~\cite{Isgur:1989vq,Luke:1990eg}
\begin{equation}
	h_{A_1}(1) = \eta_A \left[
		1_{\rm Isgur-Wise} + 0_{\rm Luke} +
		\delta_{1/m^2} + \delta_{1/m^3} \right]
\end{equation}
where $\eta_A$ is a short-distance coefficient of HQET,
and the $\delta_{1/m^n}$ contain long-distance matrix elements.
The structure of the $1/m_Q^n$ corrections is
\begin{eqnarray}
	\delta_{1/m^2} & = &
		- \frac{  \ell_V}{(2m_c)^2}
		+ \frac{ 2\ell_A}{(2m_c)(2m_b)}
		- \frac{  \ell_P}{(2m_b)^2} \label{eq:ell}\\
	\delta_{1/m^3} & = &
		- \frac{  \ell_V^{(3)}}{(2m_c)^3}
		+ \frac{  \ell_A^{(3)}\Sigma +
		\ell_D^{(3)}\Delta}{(2m_c)(2m_b)}
		- \frac{  \ell_P^{(3)}}{(2m_b)^3} \label{eq:ell3}
\end{eqnarray}
where $\Sigma=1/(2m_c)+1/(2m_b)$ and $\Delta=1/(2m_c)-1/(2m_b)$.
One must make sure to calculate $\eta_A$ and the $\ell$s in the same 
renormalization scheme.

Lattice gauge theory with Wilson fermions has the same heavy-quark
symmetries as continuum~QCD, for all~$m_Qa$.
It therefore admits a description with HQET, provided
$m_Q\gg\Lambda$~\cite{El-Khadra:1997mp,Kronfeld:2000ck,%
Harada:2001fi,Harada:2001fj}.
In this description, HQET matrix elements, such as the $\ell$s in
Eqs.~(\ref{eq:ell}) and~(\ref{eq:ell3}), are essentially the same as
for continuum QCD.
So, one needs some quantities with small statistical and normalization
errors, whose heavy-quark expansion contains the~$\ell$s.
Then, one calculates the short-distance part in perturbation 
theory~\cite{Lepage:1993xa,Brodsky:1983gc},
extracts the $\ell$s from a fit, and reconstitutes~$h_{A_1}(1)$.

In our work on the $B\to D$ form factor~\cite{Hashimoto:2000yp}, we 
found that certain ratios have the desired low level of uncertainty:
\begin{eqnarray}
	\frac{\langle D   |\bar{c}\gamma^4 b| B   \rangle
		  \langle B   |\bar{b}\gamma^4 c| D   \rangle}
		 {\langle D   |\bar{c}\gamma^4 c| D   \rangle
		  \langle B   |\bar{b}\gamma^4 b| B   \rangle} & = &
	\left\{ \eta_V^{\rm lat} \left[ 1 - \ell_P\Delta^2 -
		\ell_P^{(3)} \Delta^2\Sigma \right] \right\}^2 ,
	\label{eq:R+} \\
	\frac{\langle D^* |\bar{c}\gamma^4 b| B^* \rangle
		  \langle B^* |\bar{b}\gamma^4 c| D^* \rangle}
		 {\langle D^* |\bar{c}\gamma^4 c| D^* \rangle
		  \langle B^* |\bar{b}\gamma^4 b| B^* \rangle} & = &
	\left\{ \eta_V^{\rm lat} \left[ 1 - \ell_V \Delta^2 -
		\ell_V^{(3)} \Delta^2\Sigma \right] \right\}^2 ,
	\label{eq:R1} \\
	\frac{\langle D^* |\bar{c}\gamma^j \gamma_5 b| B   \rangle
		  \langle B^* |\bar{b}\gamma^j \gamma_5 c| D   \rangle}
		 {\langle D^* |\bar{c}\gamma^j \gamma_5 c| D   \rangle
		  \langle B^* |\bar{b}\gamma^j \gamma_5 b| B   \rangle} & = &
	\left\{ \check{\eta}_A^{\rm lat} \left[ 1 - \ell_A\Delta^2 -
		\ell_A^{(3)} \Delta^2\Sigma \right] \right\}^2 .
	\label{eq:RA1} 
\end{eqnarray}
For lattice gauge theory, the heavy-quark expansions in
Eqs.~(\ref{eq:R+})--(\ref{eq:RA1}) have been derived
in Ref.~\cite{Kronfeld:2000ck}, leaning heavily on
Refs.~\cite{Falk:1993wt}.
The one-loop expansions of $\eta_V^{\rm lat}$ and 
$\check{\eta}_A^{\rm lat}$ are in Ref.~\cite{Harada:2001fj}.
Thus, these ratios yield all terms in $\delta_{1/m^3}$ 
except~$\ell_D^{(3)}$.

We wish to obtain the $1/m_Q^2$ corrections to the double ratios, 
but the lattice action and currents do not normalize all such terms 
correctly.
HQET reveals several sources of such contributions, in a systematic 
way~\cite{Falk:1993wt,Kronfeld:2000ck}.
The most crucial are the $1/m_Q^2$ corrections to the currents, 
which enter the double ratios as follows:
\begin{equation}
	\frac{
	[1 - \lambda(X_b/m_b^2 - 1/m_cm_b + X_c/m_c^2)]^2
	}{
	[1 - \lambda(2X_c - 1)/m_c^2]
	[1 - \lambda(2X_b - 1)/m_b^2]
	} =
	1 - \lambda\left(\frac{1}{m_c} - \frac{1}{m_b}\right)^2,
\end{equation}
where $\lambda$ is proportional to $\lambda_1$ or $\lambda_2$, and
$X_Q/m_Q^2$ indicates incorrect normalization.
The correct normalization of the $1/m_cm_b$ terms is built into the 
current we used.
The cancellation of the others is a key feature of the double ratios.
Other contributions either vanish or are correctly normalized to 
order~$\alpha_s$~\cite{Kronfeld:2000ck}.
This, and other matching uncertainties of order $\alpha_s^2$ and 
$(\bar\Lambda/m_Q)^3$, are put into the error budget.

\begin{figure}[b!]
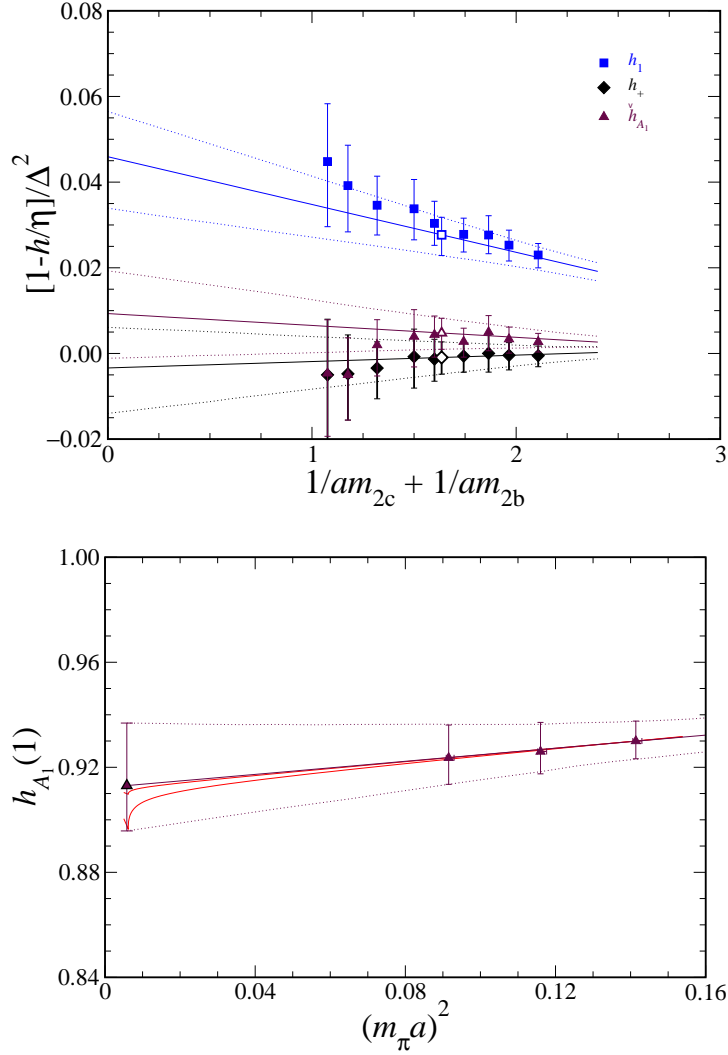

\begin{center}
	\epsfig{file=mass59.eps,width=0.60\textwidth}\\[14pt]
	\epsfig{file=h_vs_mpi2.eps,width=0.60\textwidth}
	\caption[fig:1]{(a) Heavy-quark mass dependence of the double ratios.
	(b) Chiral extrapolation of $h_{A_1}(1)$.}
	\label{fig:1}
\end{center}
\end{figure}

Fig.~\ref{fig:1}(a) shows the heavy-quark mass dependence from 
Ref.~\cite{Hashimoto:2001nb}.
As expected, $\ell_V$ is the largest of the $1/m_Q^2$ matrix elements.
Because of the fit, the value of $\ell_V$ is highly correlated with 
that of $\ell_V^{(3)}$, but the physical combination is better 
determined.

We have also studied the dependence of the calculation on the mass of 
the light spectator quark, over the range $0.4\le m_q/m_s\le 1$.
As seen in Fig.~\ref{fig:1}(b), there is a slight linear dependence on 
$m_\pi^2$, which is proportional to~$m_q$.
The points are correlated, so the trend is significant.
The main effect of extrapolating in $m_\pi^2$ is to increase the 
statistical error.
In addition, there must be a pion loop contribution~\cite{Randall:1993qg}, 
which is mistreated in the quenched approximation.
We treat the omission of this effect as a systematic error.

\begin{figure}[b!]
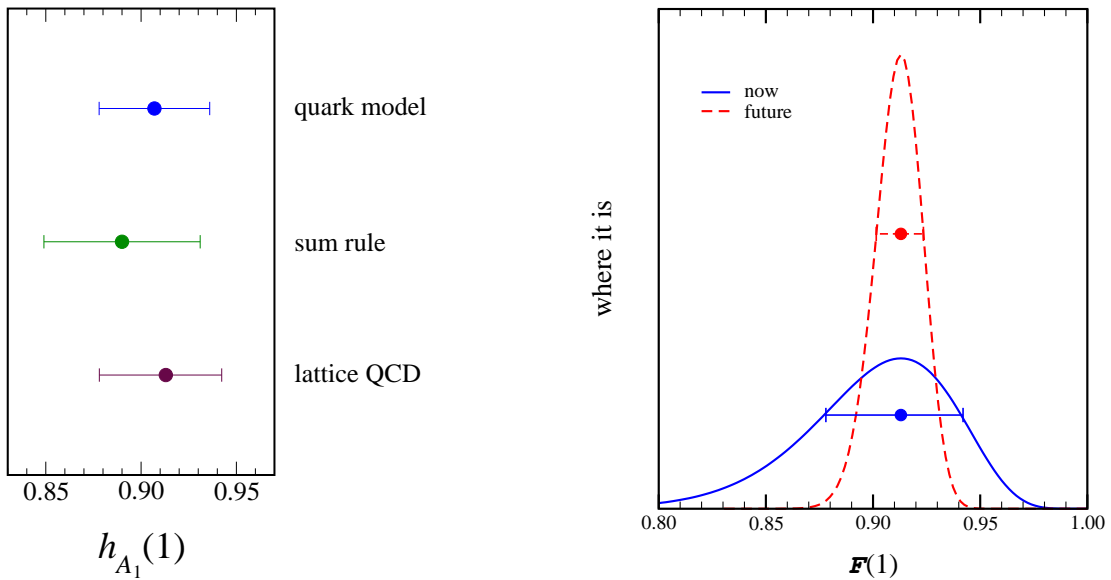

\begin{center}
	\epsfig{file=hA1.eps,height=3.0in}\hspace{0.8in}
	\epsfig{file=poisson.eps,height=3.0in}
	\caption[fig:2]{(a) Comparison of methods.
	(b) Simple Ansatz for more (and less) likely values, now and in the 
	future.}
	\label{fig:hA1}
\end{center}
\end{figure}

After putting everything back together again,
we find~\cite{Hashimoto:2001nb}
\begin{equation}
	{\cal F}_{B\to D^*}(1) = 0.913 
		{}^{+0.024}_{-0.017} 
		\pm 0.016 
		{}^{+0.003}_{-0.014} 
		{}^{+0.000}_{-0.016} 
		{}^{+0.006}_{-0.014} ,
	\label{eq:result}
\end{equation}
where the uncertainties stem, respectively, from
statistics and fitting,
HQET matching,
lattice spacing dependence,
the chiral extrapolation,
and the effect of the quenched approximation.
In Fig.~\ref{fig:hA1}(a) we compare our result for ${\cal F}_{B\to D^*}(1)$ 
against the estimate based on the quark model~\cite{Neubert:1994vy}, and 
on the sum rule~\cite{Uraltsev:2000qw}.
The defects are as follows:
The quark model omits some dynamics (more than quenching), and it is
not clear that it gives the $\ell$s in the same scheme as~$\eta_A$.
The sum rule has an incalculable contribution from excitations with
$(M-m_{D^*})^2<\mu^2$, which can only be estimated.
The present lattice result is in quenched approximation, but the 
error from quenching is the last error bar in Eq.~(\ref{eq:result}).

For using this result in a global fit to the CKM matrix, it is useful 
to have some idea what values of theoretical quantities are more (or 
less) likely.
A~flat distribution would be incorrect, because the first error in 
Eq.~(\ref{eq:result}) is essentially statistical, and the others are 
under some control.
Also, one cannot rule out a tail for lower values; they are just 
unlikely.
Finally, we know that ${\cal F}_{B\to D^*}(1)\le1$~\cite{Shifman:1995jh}.
A~simple formula that captures these features is the Poisson 
distribution
\begin{equation}
	P(x) = N x^7 e^{-7x}, \quad 
	x = \frac{1-{\cal F}_{B\to D^*}(1)}{0.087}.
\end{equation}
In the future one could reduce the uncertainty by a factor of~3,
as sketched in Fig.~\ref{fig:hA1}(b), and one could provide a 
distribution stemming from the Monte Carlo calculation, and properly
propagated through the systematic analysis.
\vfill
We thank Aida El-Khadra for helpful discussions.
High-performance computing was carried out on ACPMAPS; we thank past
and present members of Fermilab's Computing Division for designing,
building, operating, and maintaining this supercomputer, thus making
this work (ACPMAPS' last) possible.
Fermilab is operated by Universities Research Association Inc.,
under contract with the U.S.\ Department of Energy.
SH is supported in part by the Grants-in-Aid of the
Japanese Ministry of Education under contract No.~11740162.

\end{document}